\documentclass[12pt,a4paper]{article}
\usepackage[parfill]{parskip}    % Activate to begin paragraphs with an empty line rather than an indent
\usepackage[english]{babel}
\usepackage[applemac]{inputenc}
\usepackage{fullpage}
\usepackage{graphicx}
\usepackage{amssymb}
\usepackage{hyperref}
\usepackage{setspace}

\begin{document}

\doublespacing
\textbf{Comparison of the photoluminescence properties of semiconductor quantum dots and non-blinking diamond nanoparticles. Observation of the diffusion of diamond nanoparticles in living cells.}

\emph{Orestis Faklaris*, Damien Garrot, Vandana Joshi,  Jean-Paul Boudou, Thierry Sauvage, Patrick Curmi, and Fran\c cois Treussart}\\

[*] O.~Faklaris, Dr.~D.~Garrot, Prof.~F.~Treussart \\
Laboratoire de Photonique Quantique et Mol\'eculaire,\\
Ecole Normale Sup\'erieure de Cachan,
UMR CNRS 8537, Cachan (France)\\
orestis.faklaris@lpqm.ens-cachan.fr

 Dr. V.~Joshi, Dr. P.A.~Curmi\\
 Laboratoire Structure et Activit\'e des Biomol\'ecules Normales et Pathologiques, \\
 Universit\'e Evry-Val-d'Essonne and INSERM U829, Evry (France)

 Dr.~J.-P.~Boudou\\
Laboratoire LMSSMat,
Ecole Centrale Paris, 
UMR CNRS 8579, Châtenay-Malabry (France)
 
 Dr.~T.~Sauvage\\
Laboratoire CEMHTI (Conditions Extr\^emes et Mat\'eriaux : Haute Temp\'erature et Irradiation),
 UPR CNRS 3079, Orl\'eans (France)

Keywords :  quantum dots, nanodiamonds, biomarker, photoluminescence, photostability

\newpage
\textbf{Abstract}

Long-term observations of photoluminescence at the single-molecule level were until recently very difficult, due to the photobleaching of organic fluorophore molecules. Although inorganic semiconductor nanocrystals can overcome this difficulty showing very low photobleaching yield, they suffer from photoblinking. A new marker has been recently introduced, relying on diamond nanoparticles containing photoluminescent color centers. In this work we compare the photoluminescence of single quantum dots (QDs) to the one of nanodiamonds containing a single-color center. Contrary to other markers, photoluminescent nanodiamonds present a perfect photostability and no photoblinking.  At saturation of their excitation, nanodiamonds photoluminescence intensity is only three times smaller than the one of QDs. Moreover, the bright and stable photoluminescence of nanodiamonds allows wide field observations of single nanoparticles motion. We demonstrate the possibility of recording the trajectory of such single particle in culture cells.

\section{Introduction}
An important requirement to understand biomolecules interactions is to be able to probe each of them individually at the single molecule scale. For this purpose a stable and reliable marker is needed. Contrary to organic dyes and fluorophores that are widely used but photobleach after some time of illumination, semiconductor nanocrystals (or QDs) have a low photobleaching yield~\cite{Michalet_05}. In addition, QDs offer the possibility of multicolor staining by size tuning. However they may be cytotoxic on long-term scale~\cite{Kirchner_05} and they present the important drawback of showing intermittency in their emission, also known as photoblinking~\cite{Nirmal_96}. 

In contrast photoluminescent nanodiamonds (PNDs) are promising alternative biomarkers. Their photoluminescence results from embedded nitrogen-vacancy color centers (NV)~\cite{Gruber_97} in the diamond matrix, which present perfect photostability : no photobleaching, nor photoblinking. Such remarkable properties allow reliable single particle tracking~\cite{Neugart_07,Chang_08}.
The low cytotoxicity of nanodiamonds produced by High Pression and High Temperature (HPHT)~\cite{Fu07, Faklaris_08} and of nanodiamonds produced by detonation~\cite{Schrand07, Huang07} has also been demonstrated. In addition PNDs are internalized spontaneously in different cell lines, either by endocytosis or by other mechanisms~\cite{Faklaris_08}. These advantages give to PNDs the potential of multiple applications in biology, for example as a marker of different intra or extra cellular compartments or as a long term traceable delivery vehicle for biomolecules translocation in cell. 

In this work, we compare the photoluminescent properties of QDs and PNDs. We selected QDs as a reference because they are widely used and have the best photoluminescent properties among inorganic biomarkers. 
The photoluminescence properties of individual nanoparticles of both types were studied with a home-made confocal microscope equipped with single photon detectors, a time correlation measurement setup and an imaging spectrograph.

Interestingly we observed the photoblinking of QDs and showed that, on the contrary, the PNDs photoluminescence remains stable in time. We compared the photoluminescence intensity of single QDs with the one of single NV emitters and found only a three fold decrease. 
In addition, we studied internalized PNDs in living cells in culture, using wide field laser illumination microscopy. PNDs trajectories were recorded with a standard cooled CCD array. We observed a confined motion of these particles inside the cell cytoplasm, contrary to their free browanian motion measured independently in water:glycerol mixture used as a calibration medium. Such measurements show that the PNDs can be considered as reliable long-term markers for intracellular studies.

\section{Experimental Methods}

\subsection{Quantum Dots and Photoluminescent Nanodiamonds preparation}

\begin{figure}[htbp]
\begin{center}
\includegraphics[width=0.9\textwidth]{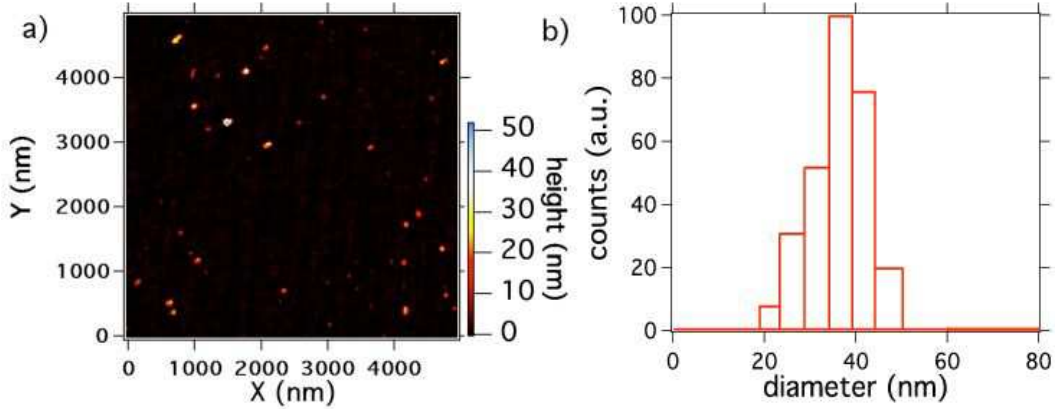}
\caption{Size distribution of the PDNs (initial nanodiamond size $<50$~nm) redispersed in the aqueous solution: a) Atomic Force Microscope scan of PNDs deposited on a mica plate (apparatus: \emph{Nanoscope IIIa}, Veeco Instruments Inc., USA ). b) Dynamic Light Scattering measurement of PNDs in the aqueous solution (apparatus: \emph{BI-200SM}, Brookhaven Instruments Corp., USA).}
\label{fig:figure1}
\end{center}
\end{figure}

The Quantum Dots used are Qdot655-IgG conjugates ( \emph{lot N$^\circ$ 51017A}, Invitrogen, USA). They are CdSe core QDs, with a shell of ZnS. This specific QD reference was chosen because its photoluminescence emission peak is centered on 655~nm (\textbf{Figure~\ref{fig:figure2}}b), close to the NV color center emission spectrum maximum (\textbf{Figure~\ref{fig:figure4}}b).
The diamond nanoparticles are turned into PNDs following the procedure described in Ref.~\cite{Yannick08}. The starting material is synthetic type Ib diamond powder (\emph{SYP 0-0.05}, Van Moppes, Switzerland) with a specified size smaller than 50~nm. The creation of the vacancies is performed by high energy proton irradiation (Van de Graaff accelerator; dose: $5\times10^{15}$~H$^+$/cm$^2$, energy: 2.6~MeV) of nanodiamonds and their stabilization next to a nitrogen impurity (forming the stable complex NV) results from subsequent annealing (800$^\circ$C, under inert gas, for 2~hours). For the irradiation process the nanodiamonds were deposited on a silicon wafer as compact thin layers of thickness smaller than 30~$\mu$m, to ensure a uniform spatial distribution of the vacancies created by the ion beam in the diamond matrices, as predicted by SRIM Monte Carlo simulations. 
After these physical processes, the deagglomeration of the nanodiamonds is achieved by strong acid treatment (H$_2$SO$_4$:HNO$_3$, 1:1 vol:vol, at 75$^\circ$C for 3~days) and 2 to 3 times rinsing by deionised water, to finally obtain a stable aqueous suspension. The zeta potential of this solution was -41.3 mV (apparatus: \emph{Zetasizer Nano ZS}, Malvern, England), a high value probably due to the abundance of carboxylic groups on the nanodiamond surface, consecutive to the strong acid treatment~\cite{Fu07}.  The PNDs size distribution, measured after their whole preparation process, yields a mean value of 41~nm (\textbf{Figure~\ref{fig:figure1}}), proving that we were able to redisperse the nanodiamonds to their primary size. By repeated centrifugations, we can further select an ensemble of smaller PNDs from this suspension.

\subsection{Confocal imaging of QDs and PNDs}

The QDs and PNDs are deposited on the glass substrates by spin-coating. The photoluminescence of the QDs and the PNDs is studied with a home-made scanning stage confocal microscope (see \emph{Supplementay File} for its detailed description). We use a cw excitation laser at a wavelength of 532~nm well adapted to the QD and NV center~\cite{Mita_96}  absorption spectra. The number of NV color centers per diamond nanoparticle is measured using photon antibunching experiments, relying on a Hanbury Brown and Twiss time intensity correlator~\cite{Yannick08,Treussart_PhysB06}.
 
 \subsection{Wide-field imaging and cells preparation}
The experimental setup was the same as the one used for confocal imaging, with two modifications.  First, the excitation 532~nm laser beam was defocused before entering the objective of the microscope by a 300 mm lens, making the beam converge onto the back focal plane of the objective. Secondly, the side output port of the microscope stand was used to image the focal plane onto a sensitive CCD array (\emph{CoolSnap Monochrome}, Photometrics, USA). 

\emph{HeLa} cells were grown in standard conditions on glass coverslips in Dulbecco's Modified Eagle Medium (DMEM) as the culture medium, supplemented with 10\% Foetal Calf Serum (FCS). To study the internalization of PNDs, cells were seeded at a density of $10^5$~cells/1.3~cm$^2$ and grown at 37$^\circ$C in a humidified incubator under 5\% CO$_2$ atmosphere. Twenty-four hours after the cell seeding, the PNDs aqueous suspensions were added to the cell medium. We performed the microscope observations after two hours incubation of the cells with the PNDs .

\section{Results and discussion}

In the following, we compare the saturation behaviour of the photoluminescence signals from single QDs and single PNDs, upon increase of the laser excitation power. We study few emitters of each type and we then extract a mean behaviour. The excitation laser beam is therefore polarized circularly to be insensitive to the different single-emitter dipole orientations, which could have led to some artifact.

\subsection{Photoluminescence study of single Quantum Dots}

\begin{figure}[htbp]
\begin{center}
\includegraphics[width=0.9\textwidth]{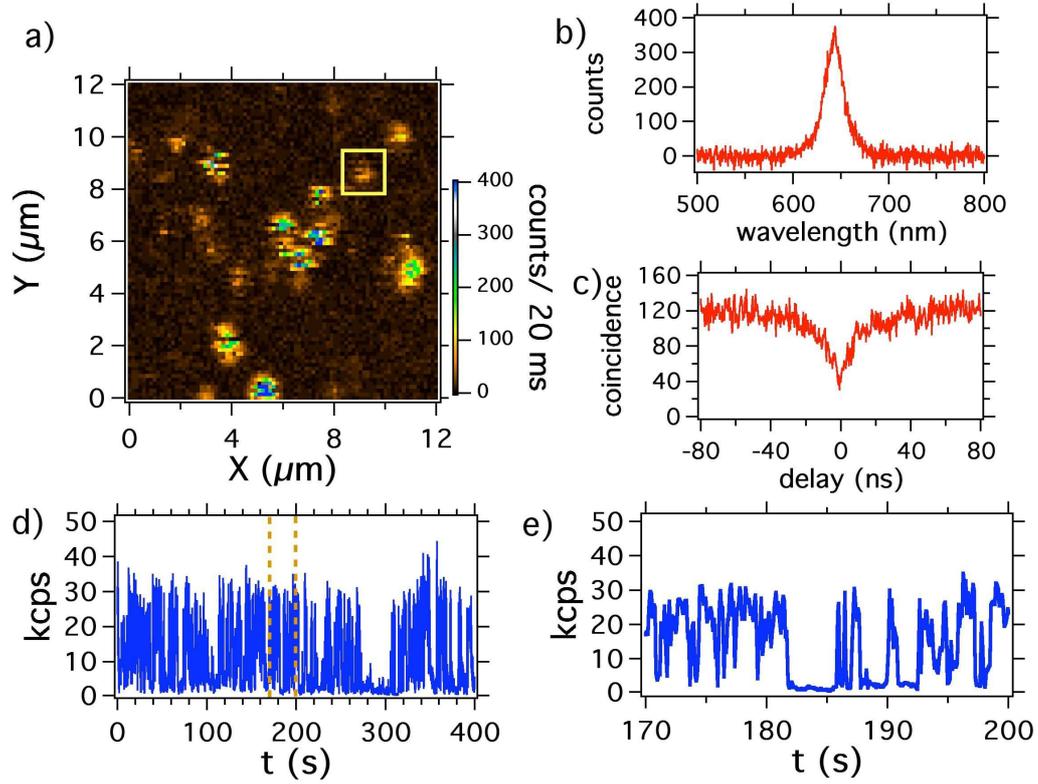}
\caption{Photoluminescence intensity of single QDs showing photoblinking. The cw excitation laser used has a wavelength of 532~nm. a) Confocal raster scan of a sample made of QDs spincoated on a glass coverslip, laser excitation power: $5~\mu$W; the blinking of QDs during the scan acquisition can be observed as dark lines interrupting bright spots. b) Photoluminescence spectrum of the QD surrounded with a yellow square on the raster scan a). c) Intensity time correlation histogram of the same QD, showing antibunching at zero delay which proves that we address a single emitter; d) A 400~s long photoluminescence intensity trajectory of the same single QD (binning: 100~ms), laser excitaton power: $10~\mu$W; e) zoom of the photoluminescence time-trace of d).}
\label{fig:figure2}
\end{center}
\end{figure}

\textbf{Figure~\ref{fig:figure2}a} shows a confocal raster scan of QDs deposited on a glass coverslide. Most of the isolated spots are well separated from each other except one aggregate in the center of the image.  The characteristic peak around 655~nm in the photoluminescence spectrum from a single spot allows us to confirm that the emission comes from a QD. The intensity time-correlation measurement of the light emitted from each isolated spot yields a photon antibunching corresponding to a single QD. 
To study the single-particle photoluminescence, we record the intensity during a ``long" integration time of 400~s. We clearly observe blinking in the photoluminescence, corresponding to the emitter switching randomly between an ``on" and an ``off" state (\textbf{Figure~\ref{fig:figure2}d}).

Considering the large fluctuations of the QDs photoluminescence intensity due to this blinking, defining a photoluminescent intensity is somewhat arbitrary (see \textbf{Figure~\ref{fig:figure2}e}, for example) and depends on whether we consider that the emission obeys a simple two states ``on''-``off'' model, or rather a three-state or even a continuous-state model~\cite{Verberk_02, Zhang_06}.
In most of the reported works the methods that are used to quantify the photoluminescence intensity signal introduce a threshold level below which the QDs are considered to be in the ``off'' state. 
Despite the fact that more complicated techniques have recently been developed, like the changepoint method \cite{Watkins_05}, we decided to use a threshold limit and set it at 50\% of the maximum photoluminescence intensity observed on the time window of the measurement.

\begin{figure}[htbp]
\begin{center}
\includegraphics[width=0.9\textwidth]{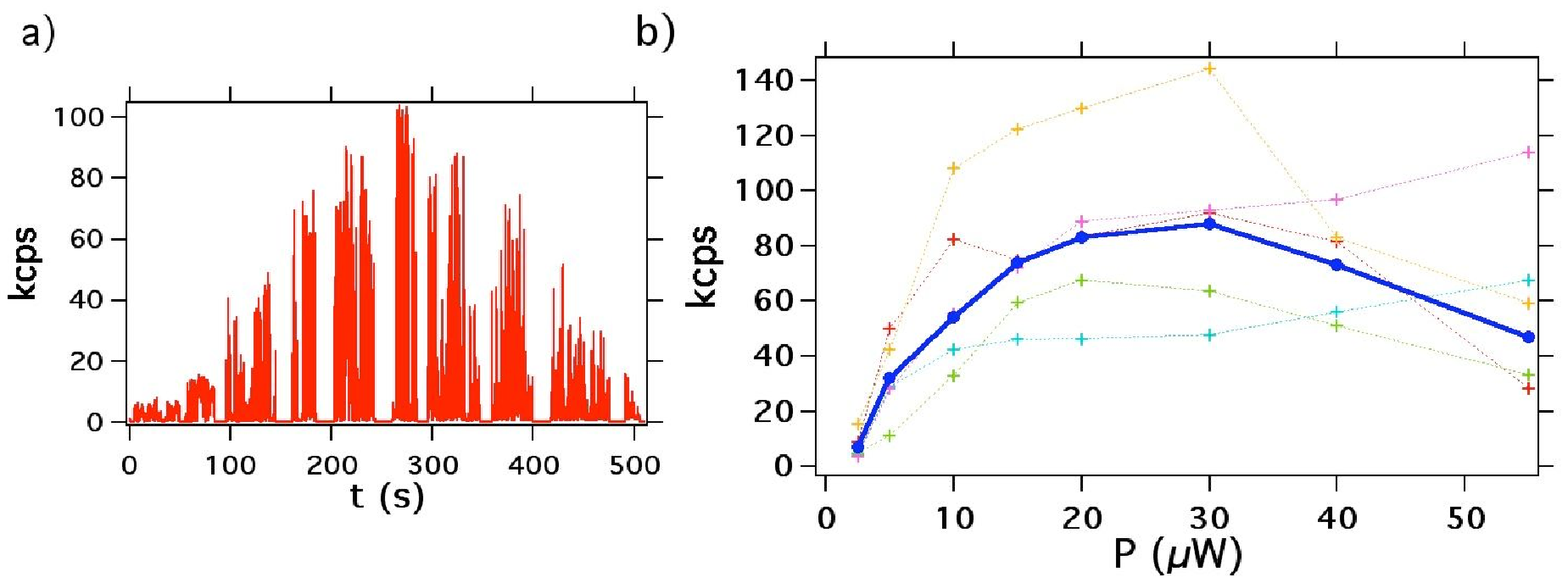}
\caption{Photoluminescence intensity vs laser excitation power for QDs. a) Single QD photoluminescence intensity dependence with excitation power, consecutive values equal to $2.5~\mu$W, $5~\mu$W, $10~\mu$W, $12~\mu$W, $15~\mu$W, $20~\mu$W, $30~\mu$W, $40~\mu$W, $55~\mu$W and $100~\mu$W. For every change of the excitation laser power, we shut off the beam for a few seconds in order to measure exactly its the power value; binning time: 50~ms. b) Mean saturation curve (plain blue) calculated from the average of 5 single QDs saturation curves (dashed colored).}
\label{fig:figure3}
\end{center}
\end{figure}

\textbf{Figure~\ref{fig:figure3}a} shows the photoluminescence intensity vs the laser excitation power for a single QD. Most of the QDs studied show a decrease of their photoluminescence intensity at a high excitation power, which almost vanishes at intensities higher than 100~kW/cm$^2$ (corresponding to 100~$\mu$W input power). We tentatively attribute this effect to photoinduced oxidation of the QDs surface. The photoluminescence saturation behaviour is recorded for 5 single QDs from which an average curve is inferred (see \textbf{Figure~\ref{fig:figure3}b}). The later indicates that the saturation is achieved for $20~\mu$W excitation power corresponding to a maximum photoluminescence intensity of 82 kcounts/s (kcps). This value is compared with the one found for PNDs in the following.

\subsection{Photoluminescence study of single NV color centers in nanodiamonds}

\subsubsection{Results}
In order to compare the photoluminescence of QDs with that of PNDs we performed similar saturation measurements for single NV color centers in the diamond nanocrystals of mean size 41~nm.
\textbf{Figure~\ref{fig:figure4}a} shows a confocal raster scan of PNDs deposited on a glass coverslip. 
The photoluminescence spots observed have emission intensities varying a lot from one spot to the other. This is due to an inhomogeneous number of NV centers per nanocrystal. 
We selected a nanocrystal containing a single NV center (assessed by \textbf{Figure~\ref{fig:figure4}c} antibunching measurement) and we recorded its saturation curve. The saturation occurs for excitation power value higher than 1.5~mW, corresponding to a photoluminescence intensity of 25~kcps. Note that compared to QDs, the NV photoluminescence intensity is constant in time (\textbf{Figure~\ref{fig:figure5}a}).
The Signal-to-Noise Ratio (SNR) for this PND is about 10 (ratio of $\langle N\rangle=380$ to the standard deviation $\sigma=37$), while the shot-noise-limited SNR is by definition $\sqrt N\approx 19.5$. 
This excess of noise can be explained by a bunching effect in the photon stream that shows up as a $g^{(2)}(\tau)$ greater than 1 ({Figure~\ref{fig:figure4}c}).

\subsubsection{Discussion}
According to our measurements, a single QD yields an emission intensity at saturation three times larger than the one of a single NV center. In addition the excitation laser saturation power for a single QD is $\times 75$ lower than the one required for a NV center. This result can be explained by the difference in the absorption cross section $\sigma$ of the two type of emitters. 
The absorption cross-section was inferred in the range $\sigma_{\rm QD}\approx 2-16\times 10^{-16}$~cm$^2$~\cite{Moerner_CPL00} at 488~nm excitation wavelength for the QD, while for the NV center $\sigma_{\rm NV}\approx 3\times 10^{-17}$~cm$^2$~\cite{Chang_JPCA07}. 
Considering that the excitation intensity at saturation is proportional to $(\sigma\tau)^{-1}$ ($\tau$ being the emitter radiative lifetime), and taking into account that the QD and the NV color center have similar lifetime $\tau\approx 20$~ns, the change in the saturation intensity results only from the change in the absorption cross-sections. A maximal value of $\sigma_{\rm QD}^{\rm max}/\sigma_{\rm NV}=54$ is then expected, which is slightly lower than the one observed experimentally. This discrepancy may be due to the very small number of emitters studied which leads to non statistically significant measurements. 

Regarding the stability of the photoluminescence intensity in time, the QDs clearly show intermittent emission while the NV centers photoluminescence intensity is perfectly stable. Recent studies show that the QD core-shell can be optimized to reduce the blinking~\cite{Mahler_08, Chen_08}.  But despite the progress in suppressing the photoblinking, a significant fraction of the QDs nanocrystals still blinks. Moreover, the blinking properties depend on the excitation power and obey to non-classical statistics~\cite{Didier_05}.

\begin{figure}[htbp]
\begin{center}
\includegraphics[width=0.9\textwidth]{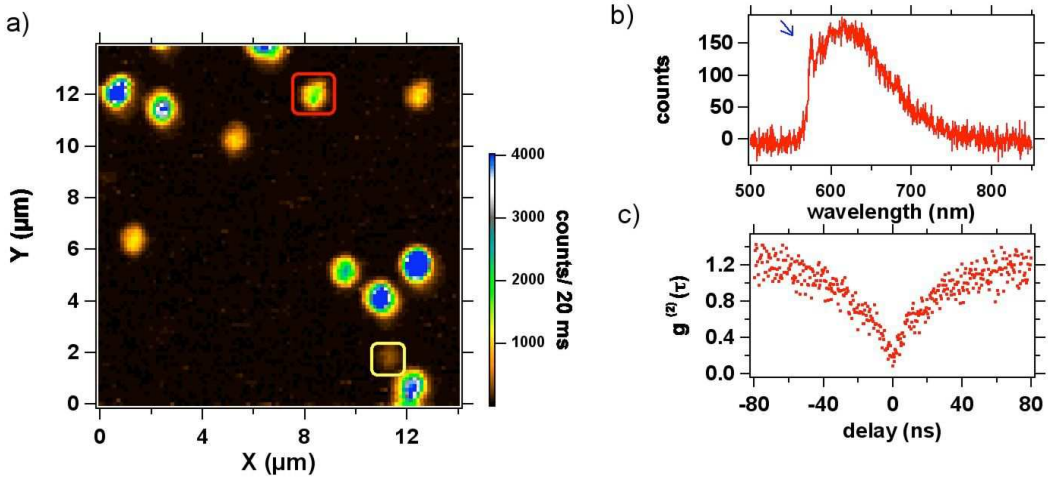}
\caption{Photoluminescence of a single PND : a) Confocal raster scan, excitation laser power: 1~mW; the well isolated spots correspond to PNDs. b) Photoluminescence spectrum of the PND surrounded in yellow (bottom right angle), corresponding to the emission of a neutral NV$^0$ color center, identified thanks to its Zero Phonon Line (ZPL, blue arrow) at 575~nm. c) Intensity time-correlation function $g^{(2)}$ of the same PND, showing photon antibunching at zero delay, associated to the emission of single photons. This observation proves that we address a single color center.}
\label{fig:figure4}
\end{center}
\end{figure}

\begin{figure}[htbp]
\begin{center}
\includegraphics[width=0.9\textwidth]{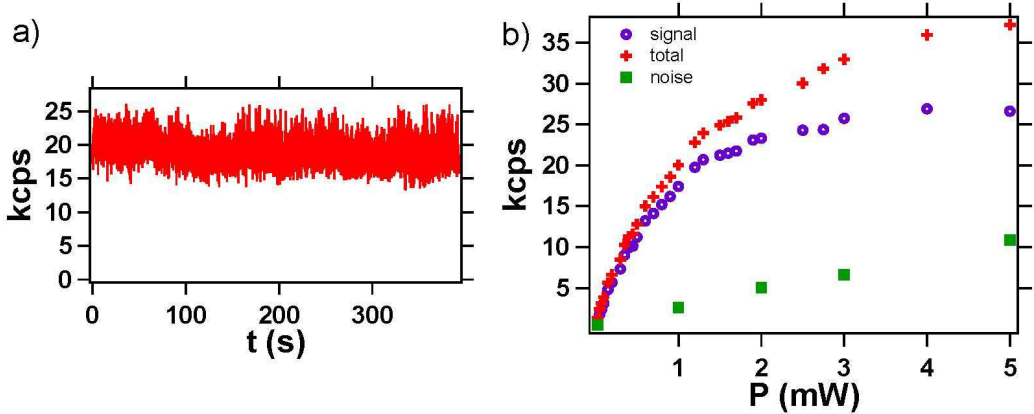}
\caption{Photoluminescence intensity stability and saturation curve for a single NV color center in a PND : a) Photoluminescence signal  (binning 20~ms) recorded over a time duration of 395~s, showing the perfect photostability of NV center emission. b) Photoluminescence intensity curve vs laser excitation power, for the single NV center surrounded in yellow in \textbf{Figure~\ref{fig:figure4}a}. Green squares correspond to the background coming from the substrate and recorded from a position a few hundreds of nanometers next to the PND; red crosses are associated to the total signal collected from the PND location; purple circles is the signal corrected from the backround.}
\label{fig:figure5}
\end{center}
\end{figure}

Even though the photoluminescence intensity from a single NV center is three times smaller than the one from a QD, if we consider their perfect photostability they are still very interesting markers for applications in biology. In addition, by selecting nanodiamonds rich in nitrogen (nitrogen concentrations larger than 100~ppm) and by optimizing the conditions for the NV color center creation (irradiation and annealing process), we managed to obtain nanocrystals of about 40~nm in size, containing up to 4-5 NV centers. On the \textbf{Figure~\ref{fig:figure4}a} raster scan, the spot circled in red corresponds to a nanocrystal containing 5 NV centers. Such a nanocrystal displays a higher photoluminescence intensity than the one of a single QD of similar size. Moreover, the irradiation dose used to create the vacancies in the PNDs can still be increased by at least one order of magnitude, using the same accelerator. Therefore, we should be able to produce even brighter PNDs.

\subsection{Wide-field microscopy observations of freely diffusing PNDs} 

For biological applications wide-field microscopy is of great importance. In such experiments, QDs have already been used as biomarkers to study the dynamics of biomolecules~\cite{Dahan_03, Cognet_07}. However due to their blinking, the tracking of a single QD appears to be tedious, which strongly limits its use in such applications.
In comparison, a PND containing many NV centers can be as bright as a QD, with the additional property of a perfectly stable emission, well suited for tracking over a long observation period.

In order to prove this statement, we first examined the free brownian motion of PNDs in solution. We used PNDs of two different sizes, the previously mentioned 41~nm and other ones of size 163~nm, as measured by dynamic light scattering (Figure S2 of the \emph{Supplementary File}).
\textbf{Figure~\ref{fig:figure6}a} is associated to the video of 163~nm in size PNDs, freely diffusing in a 20\% water - 80\% glycerol solution.
The PNDs diffusion in this solution is a free 3D-brownian motion, but the observations are restricted to the portion of this motion taking place in the focus  plane. More precisely only PNDs which are moving within a slice of thickness equal to the depth of focus ($\approx 810$~nm) are observed. Therefore the motion recorded is a 2D-projection of \emph{portions} of the 3D trajectory contained in this slice. 
To construct the particles trajectory from this record, we used the ``ParticleTracker'' plugin of NIH-\emph{ImageJ} software, which implements the algorithm of Ref.\cite{Koumou_05}.
The particles that are too far from the middle plane of the focusing slice appear as non diffraction-limited dimmer spots that are excluded with the software, using a filtering procedure. This procedure relies on a size restriction and intensity level cutoff. It filters the particles motions which take place in a slice thinner than the one limited by the optical depth of focus, so that the motion really taken into account for the trajectory construction is very close to a 2D one.

The``ParticleTracker'' plugin also constructs a trajectory with a step resolution better than the real-space pixel size of $\approx 140$~nm. To do so, it extracts the PND position in each frame from the maximum of a gaussian fit of the single emitter spot (which covers about 4 pixels). The program uses a discretization step of  23~nm which is 6 times smaller than the pixel size. This procedure allows us to infer displacements smaller than the pixel size.
The noise on the localization of a spot was also estimated of the order of 25-30~nm from measurements done on ﬁxed PNDs embedded in a polymer layer. We provide in the \emph{Supplementary ﬁle} the example of a PND trajectory inside the same cell (Figure S3c) showing a directed motion on a short distance, with a lateral spatial broadening of the same 25-30~nm order of magnitude. This observation conﬁrms the static noise measurements.

After the video processing, and the trajectory reconstruction, the PNDs were mostly found to follow brownian trajectories with no directional force. A usual method to provide a diffusion coefficient is to calculate the mean square displacements (MSD)~\cite{Saxton_97}. For spherical particles executing a 2D free brownian motion, MSD$(t) = 4Dt$, where $D$ is the diffusion coefficient and $t$ the diffusion time. In case there are external forces (drifts) acting on the particles, the equation becomes MSD$(t) = 4Dt +  v^2(t)$, where $v$ represents the velocity of the directional movement~\cite{Kusumi_93}. 
For the PNDs of size 163~nm, the diffusion coefficient measured $D=0.04~\mu$m$^2$/s. This is the mean value out of 6 particles, with lower and higher value at $D=0.01~\mu$m$^2$/s and $D=0.05~\mu$m$^2$/s respectively. A lower bound to the error bars then results from this two extreme slopes of the MSD$(t)$.
We repeat the experiment for the 41~nm mean size PNDs (\textbf{Figure~\ref{fig:figure6}d-f}). The diffusion coefficient measured is equal to $D=0.20~\mu$m$^2$/s, which is 5 times bigger than the one of 163~nm PNDs. This measurement is the mean value over 5 particle trajectories, with extremes at $D=0.05~\mu$m$^2$/s and $D=0.35~\mu$m$^2$/s as lower and higher value respectively.

The above experimental values can be compared to the theoritical predictions given by the Stokes-Einstein equation $D=kT/(6\pi\eta a)$, $a$ being the particle radius, $\eta$ the viscosity of the medium (60~cP in our case), $k$ the Boltzman constant, and $T$ the temperature. The diffusion coefficients for 41 and 163~nm  particles inferred from this equation are equal to $D=0.18~\mu$m$^2$/s and $D=0.04~\mu$m$^2$/s respectively. These values agree well with the experimental ones.

Note that the error bars on the measurement of the diffusion coefficients are large, which is an observation already reported in previous studies~\cite{Kusumi_93}. This broad distribution is mainly due to the fact that the trajectory analysis is done on a too small number of steps. The precision on the measurement of $D$ improves with the increase of the statistical sampling of the trajectory. However in a 2D study, it is not easy to record a continuous trajectory over a long observation time, since the particles, especially the small ones, are out of focus as soon as they are moving along the $z$ axis perpendicular to the focus plane.
In addition, one other reason for a broad distribution of the diffusion coefficient is the particle size distribution, which is itself also very broad (see Figure S2).

\begin{figure}[htbp]
\begin{center}
\includegraphics[width=0.9\textwidth]{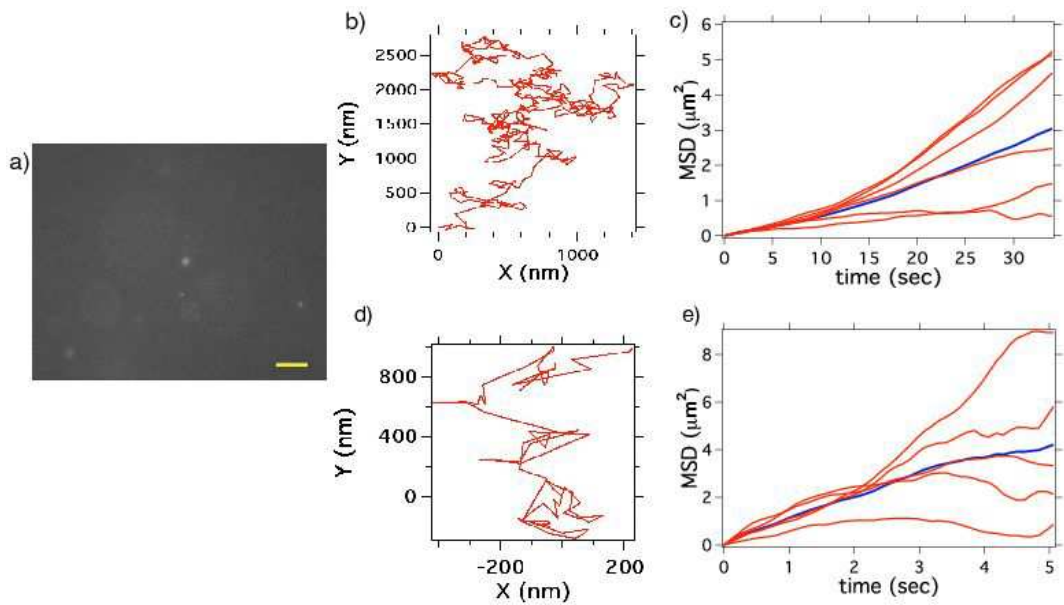}
\caption{Free brownian diffusion of PNDs in a water:glycerol mixture : a) Video sequence of 163~nm PNDs diffusing in a 20\% water - 80\% glycerol solution; exposure time per frame: 100~ms; this relatively long integration time is not a problem in our study as the water-glycerol proportion was deliberately chosen to have slow nanodiamonds motions, allowing the resolution of individual steps in time; scale bar: 5~$\mu$m.
b) Trajectory of a single PND executing a free brownian motion. c) Mean square displacements (MSD) for six different PNDs; the blue line is the mean value; d-e) similar to b-c) but for 41~nm mean size PNDs.}
\label{fig:figure6}
\end{center}
\end{figure}

\subsection{Diffusion of PNDs in living cells}

The tracking of individual biomolecules in cells is of great importance in biology. In case the biomolecule is not photoluminescent, a photoluminescent marker can be attached to it so as to follow its motion. 
If we want to use PNDs as markers, the first step is to verify that such particle can enter the cell which was the topic of a previous work~\cite{Faklaris_08} and then to follow its trajectory in time.
 
\begin{figure}[htbp]
\begin{center}
\includegraphics[width=0.9\textwidth]{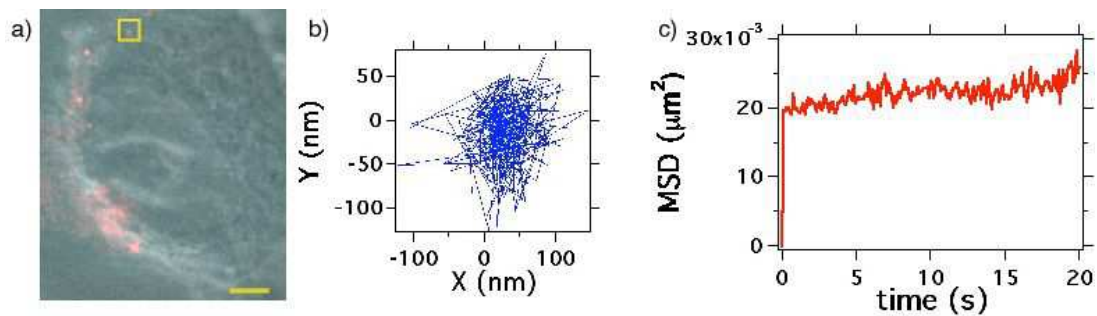}
\caption{Intracellular trafficking of single 41~nm PNDs in a live HeLa cell : a) Phase contrast image of the cell (centered on the cell nucleus) merged with fluorescence image of PNDs (in red); bar scale: 3$\mu$m. The real-time video sequence of PNDs in the cell associated to image a) is recorded with an acquisition time per frame of 100~ms. b) Trajectory of the PND located inside the yellow square. c) Mean square displacement of this PND.}
\label{fig:figure7}
\end{center}
\end{figure}

\textbf{Figure~\ref{fig:figure7}a} shows the video sequence of real-time motions of 41~nm PNDs in cell. It is clear that the motion is much more confined than in the glycerol-water solution. Note that the confinement dimension is about 160 nm, a value of about one order of magnitude larger than the noise limit. This result proves that we observed a true -although very small- motion of the PND. By analysing the MSD values at short times~\cite{Jakobson_02} we determine an equivalent ``diffusion coefficient'' $D=0.006~\mu$m$^2$/s, and among all the particle motion studied, the maximum diffusion coefficient found does not exceed the value of 0.01~$\mu$m$^2$/s, in aggreement wih previous calculations~\cite{Neugart_07}. 
The reason for the PNDs low mobility in cells is that the observed PNDs are probably captured in endosomal or lysosomal vesicles~\cite{Faklaris_08}.

\section{Conclusion}

In conclusion, we showed that in terms of photoluminescence, 41~nm PNDs can be as bright as QDs, even brighter when they contain more than 4-5 NV color centers. Contrary to QDs, their emission is perfectly stable in time. An application of such PNDs with many NV centers in their matrix is the single biomolecule tracking. As a proof of concept, we first showed that the free diffusion of single PNDs of different size can be recorded in wide field microscopy using a standard cooled CCD array detector. We further incubated PNDs with cells and observed the motion of the internalized PNDs, that appears to be confined probably in cell compartments. Our results indicate that PNDs can be used for long-term single particle tracking, and have therefore the potential of being reliable biomarkers.

\subsubsection*{Acknowledgements}
We are grateful to Jean-François Roch for fruitful discussions, and to Karen Perronet (Laboratoire Charles Fabry, Institut d'Optique, France) for the loan of the CCD array. This work was supported by the European Commission through the project ``Nano4Drugs" (contract LSHB-2005-CT-019102), by Agence Nationale de la Recherche through the project ``NaDia" (contract ANR-2007-PNANO-045), and by ``Ile de France'' Region \emph{C'Nano} grant under the project ``Biodiam".

%==============================================================
% Bibliographie
%==============================================================

\end{document}

\section*{Supporting Information}
``Comparison of photoluminescence properties of semiconductor quantum dots and non-blinking diamond nanoparticles. Observation of the diffusion of diamond nanoparticles in living cells.'',
O.~Faklaris et al.

\subsection*{Optical setup}
\begin{figure}[H]
\centerline{\includegraphics[width=13cm]{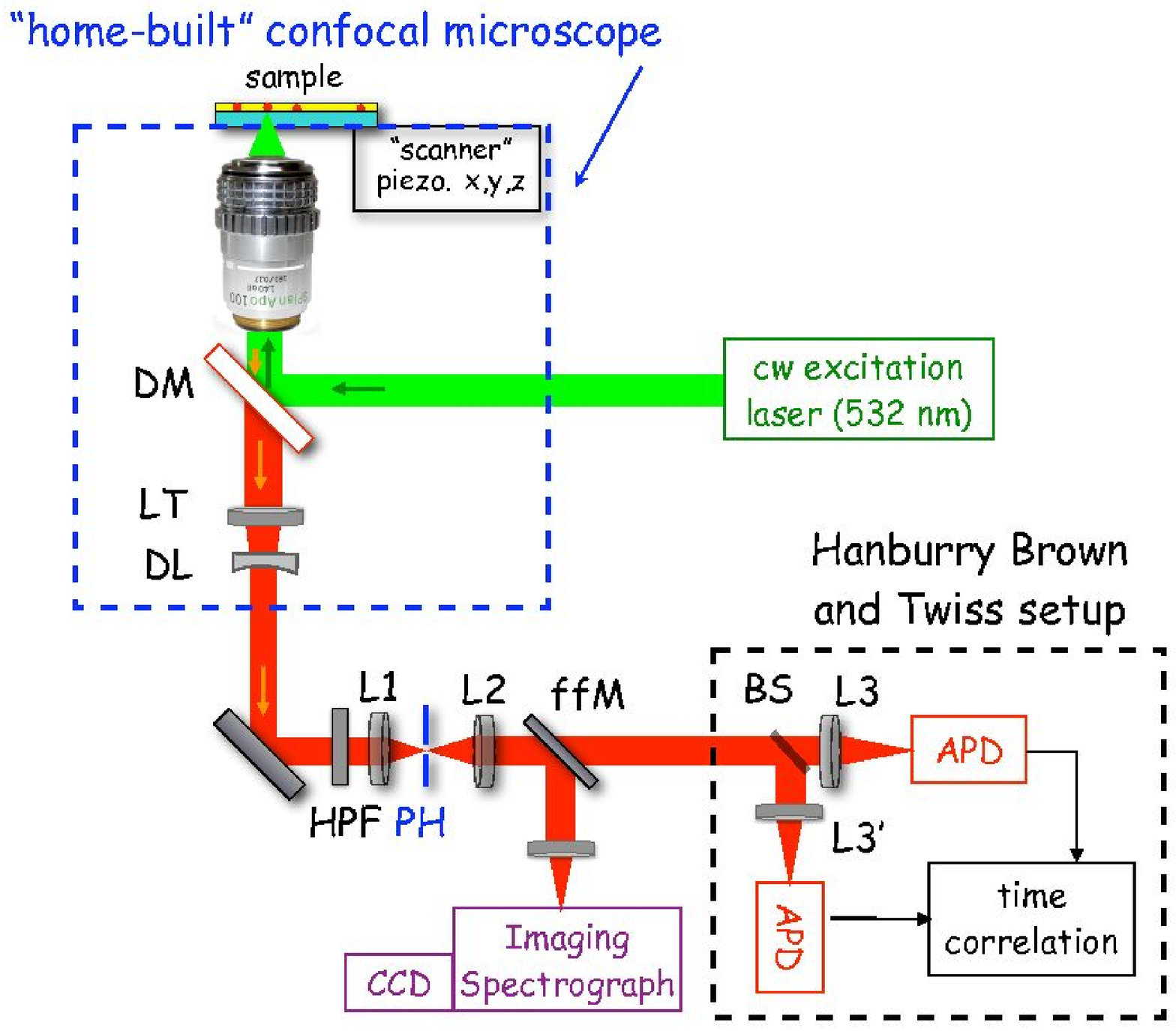}}
\caption{Optical setup used to study single-particle photoluminescence. Surrounded by the blue dashed-line box: the home-made confocal microscope. The Hanburry Brown an Twiss time correlation system is depicted in the black dashed-line box.}
\label{fig:figure1Sup}
\end{figure}

The {\bf experimental setup} used to study the single-nanoparticle (QDs and PNDs) photoluminescence analysis is depicted on {\bf Figure~S\ref{fig:figure1Sup}}.

The {\bf home-made confocal microscope}, relies on a Nikon TE300 stand, equipped with a closed loop piezoelectric stage three-axis scanner (\emph{Tritor 102}, Piezojena, Germany). After its reﬂection on the dichroic mirror DM (\emph{DCLP 540}, Chroma Corp., USA), the excitation laser beam (cw 532~nm) is focused by an oil immersion objective (Nikon Apochroma, $\times 60$, NA=1.4) onto the sample. The fluorescence is collected by the same objective, making a collimated beam going through the 200~mm focal length microscope lens tube (LT). 

Considering the microscope objective specifications, the Airy disk diameter (distance between the two first minima) in the microscope imaging plane (at the bottom of the stand) is equal to 30~$\mu$m.
However, we modified the microscope in order to get a collimated beam at its output port. This modification consists in the addition of a diverging lens DL (focal length: -125~mm)  after the lens tube turning the beam back into a collimated one, easier to handle for further propagation. The collimated output beam is then focused by the L1 lens (100~mm focal length) into the pinhole (PH). The addition of these two lenses (DL and L1) results in an increase in Airy disk size from 30 to 50~$\mu$m which is why we selected a 50~$\mu$m diameter pinhole.

The residual excitation laser light is removed with the high-pass filter HPF having a transmission of 97\% between 539-1200~nm (\emph{RazorEdge LP03-532RU-25}, Semrock, USA). 

The photoluminescence spectrum is acquired with the addition (beam intercepted with a flip-flop Mirror (ffM)) of an imaging spectrograph, relying on a simple concave grating (30\% maximum efficiency in its first diffraction order) coupled to a cooled CCD array (back-illuminated array \emph{DU-420-BV}, Andor Technology, Ireland).

The {\bf Hanbury Brown and Twiss} time-intensity correlation setup consist in two avalanche photodiodes (APD; \emph{SPCM-AQR14}, Perkin-Elmer, Canada)  in the single-photon counting mode, on both side of a non-polarizing 50/50 beam-splitter (BS), connected to a time-correlation electronics: a Time to Amplitude Converter (\emph{Ortec Model 566}, Ametek Inc., USA) with its a-output linked to a multichannel analyzer (\emph{Ortec  926-M32-USB}, Ametek Inc., USA).

\subsection*{Supplementary measurements}
\begin{figure}[!h]
\centerline{\includegraphics[width=\textwidth]{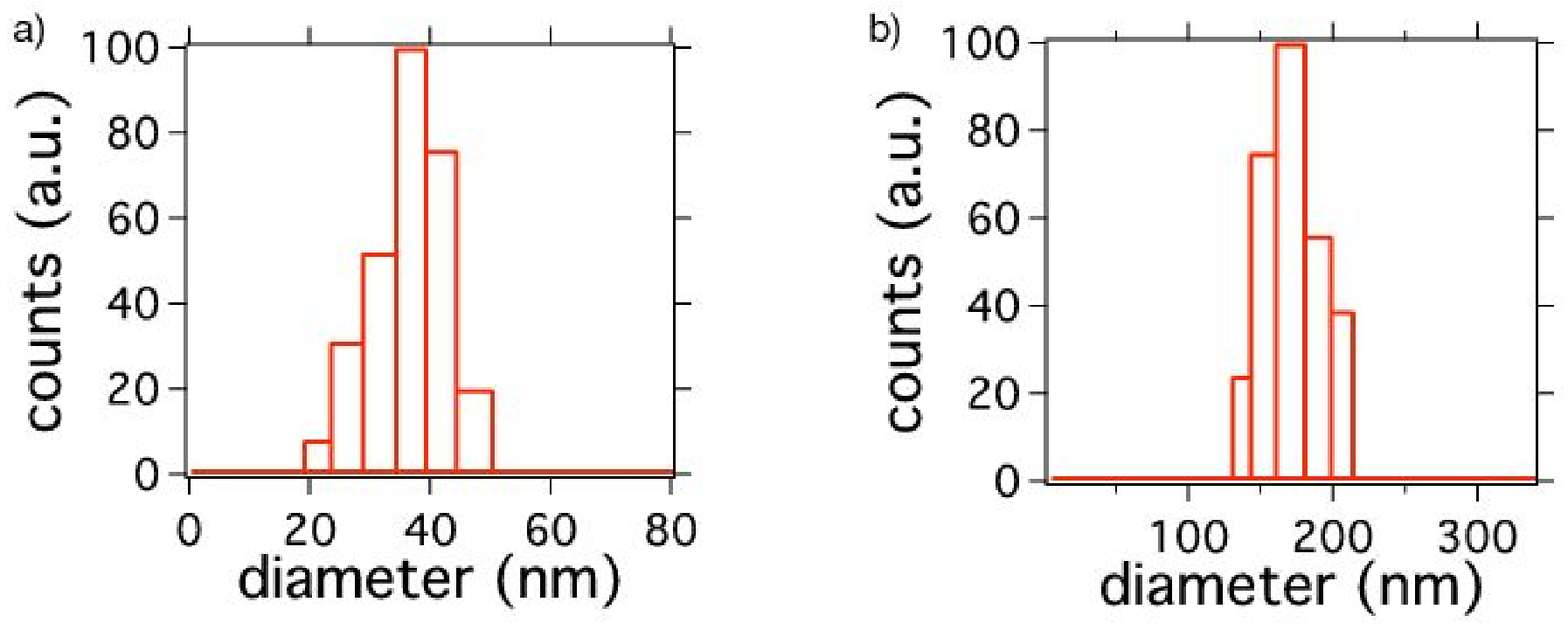}}
\caption{Size characterization of PNDs using Dynamic Light Scattering (apparatus: \emph{BI-200SM}, Brookhaven Instruments Corp., USA).
a) PNDs of 41~nm mean size. b) PNDs with a 163~nm mean size.}
\label{fig:figure2Sup}
\end{figure}

The {\bf size distribution measurements} of the two different nanodiamond types studied by video microscopy are shown of {\bf Figure~S\ref{fig:figure2Sup}}.

\begin{figure}[!h]
\centerline{\includegraphics[width=\textwidth]{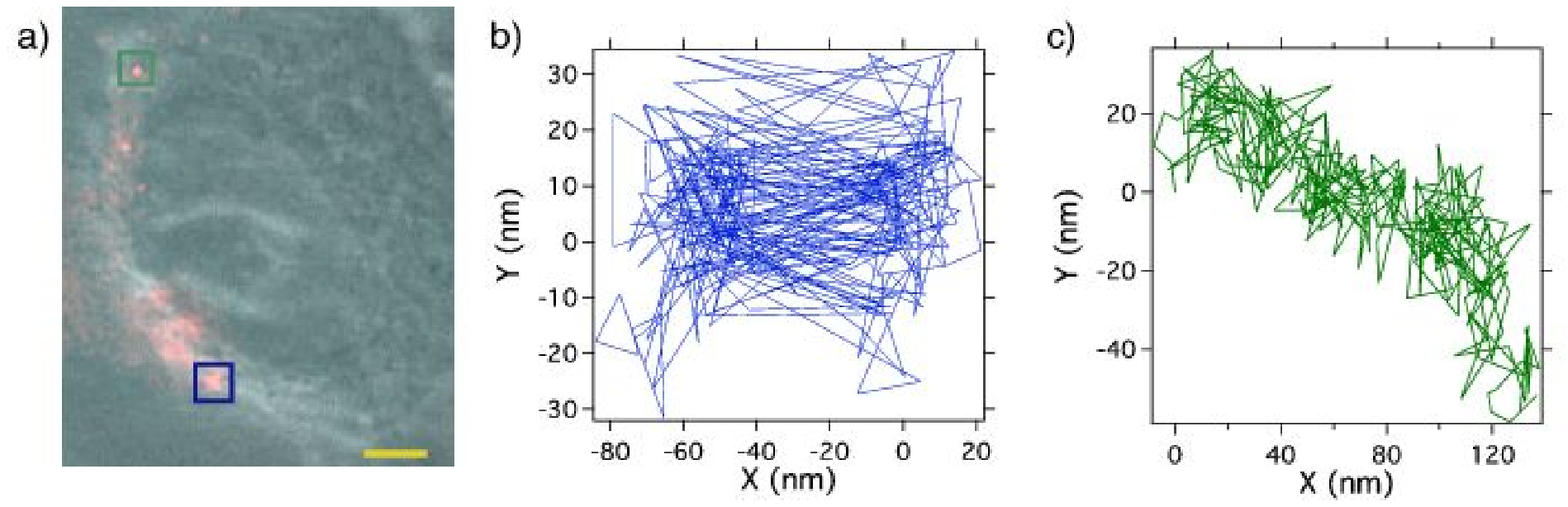}}
\caption{Trajectories of internalized nanodiamonds in the same cell as the one presented in the main text. a) Phase contrast image of cell merged with the ﬂuorescence image of PNDs (in red); scale bar 3$~\mu$m.  b) Highly conﬁned trajectory of the PND surrounded by the blue square in the image a). c) Trajectory of the PND surrounded by the green square.}
\label{fig:figure3Sup}
\end{figure}
{\bf Figure~S\ref{fig:figure3Sup}} shows two additional examples of trajectories of internalized PNDs in the cell. The {\bf Fig.~S\ref{fig:figure3Sup}b} shows a PND having a highly confined motion. It is probably an aggregate, since its fluorescence spot is not diffraction limited. {\bf Fig.~S\ref{fig:figure3Sup}c} displays the less confined trajectory, slightly directed, in comparison to the other PND (green square). In the latter case, the lateral spatial broadening of the order of 25-30~nm corresponds to the noise on the position calculated by the \emph{Image J} ``ParticleTracker'' plugin. Finally, let us mention that we did not observe any drift of the cell during these measurements.
\null
\vspace{1cm}
\centerline{\Huge $\star\star\star$}